\documentclass[fleqn,usenatbib]{mnras}

\usepackage{lineno}
\usepackage{xcolor}
\usepackage[T1]{fontenc}

\DeclareRobustCommand{\VAN}[3]{#2}
\let\VANthebibliography\thebibliography
\def\thebibliography{\DeclareRobustCommand{\VAN}[3]{##3}\VANthebibliography}

\usepackage{graphicx}	
\usepackage{amsmath}	% Advanced maths commands
\usepackage{amssymb}	% Extra maths symbols

\title[]{A Geometric Calibration of the Tip of the Red Giant Branch in the Milky Way using Gaia DR3}

\author[Dixon et al.]{M.~Dixon$^1$,
J.~Mould$^1$,
C.~Flynn$^1$,
E.~N.~Taylor$^1$,
C.~Lidman$^2$,
A.~R.~Duffy$^1$
\\
$^{1}$Centre for Astrophysics and Supercomputing, Swinburne University of Technology, Melbourne, Victoria 3122, Australia\\
$^{2}$Research School of Astronomy and Astrophysics, Australian National University, Canberra, ACT 0200, Australia\\
}

\date{Accepted XXX. Received YYY; in original form ZZZ}

\pubyear{2023}

\begin{document}
\label{firstpage}
\pagerange{\pageref{firstpage}--\pageref{lastpage}}
\maketitle

\begin{abstract}
We use the latest parallaxes measurements from Gaia DR3 to obtain a geometric calibration of the tip of the red giant branch (TRGB) in Cousins $I$ magnitudes as a standard candle for cosmology. We utilise the following surveys: SkyMapper DR3, APASS DR9, ATLAS Refcat2, and Gaia DR3 synthetic photometry to obtain multiple zero-point calibrations of the TRGB magnitude, $M_{I}^{TRGB}$. Our sample contains Milky Way halo stars at high galactic latitudes ($|b| > 36$) where the impact of metallicity, dust, and crowding are minimised. The magnitude of the TRGB is identified using Sobel edge detection, but this approach introduced a systematic offset. To address this issue, we utilised simulations with PARSEC isochrones and showed how to calibrate and remove this bias. Applying our method within the colour range where the slope of the TRGB is relatively flat for metal-poor halo stars (1.55 $<$ $(BP-RP)$ $<$ 2.25), we find a weighted average $M_{I}^{TRGB} = -4.042 \pm 0.041$ (stat) $\pm0.031$ (sys) mag. A geometric calibration of the Milky Way TRGB has the benefit of being independent of other distance indicators and will help probe systematics in the local distance ladder, leading to improved measurements of the Hubble constant.

\end{abstract}

\begin{keywords}
parallaxes, cosmology: distance scale, stars: distances
\end{keywords}

%%%%%%%%%%%%%%%%%%%%%%%%%%%%%%%%%%%%%%%%%%%%%%%%%%

%%%%%%%%%%%%%%%%% BODY OF PAPER %%%%%%%%%%%%%%%%%%

\section{Introduction}

The tip of the red giant branch (TRGB) is a widely used standard candle that can be used to measure distances up to 10 Mpc with $\sim$5$\%$ precision. This is the essential first step of the astronomical distance ladder which requires a zero-point calibration of standard candles, commonly achieved using Cepheid variable stars. This forms a path to measuring the absolute magnitude of type Ia supernovae (SNe Ia) in nearby anchor galaxies, and then to measure the Hubble constant ($H_{0}$) \citep{Dhawan_2020,Khetan2020, Freedman_2021, Riess_2022}. 

The present state of cosmology is facing a major crisis known as the Hubble tension, where there is a fundamental disagreement in the measurements of $H_{0}$, with a tension of around $5-6$$\sigma$ in $H_{0}$ \citep[see a review in][]{Di_Valentino_2021}.
The most significant discrepancy is between the values derived from modeling the cosmic microwave background (CMB) \citep{Planck2018} and the local distance ladder method that uses Cepheids \citep{Riess_2022}.

The TRGB is an alternative to the Cepheids and importantly allows $H_{0}$ to be measured independently of other distance indicators, eliminating one source of systematic uncertainty in $H_{0}$. \cite{Freedman_2021} provides the most significant measurement of $H_{0}$ with the TRGB, utilising SNe Ia host galaxies from the Carnegie Supernova Project and found a value comparable to that of CMB observations. However, there are other TRGB studies that have produced higher values of $H_{0}$, which are closer to the Cepheid value and depend on the calibrator and combination of other techniques (\citealp{Yuan_2019, Reid_2019, Dhawan_2022, Anand_2022, Anderson_2023}). 

The TRGB is a stage in the stellar evolution of low-mass ($< 2M_{\bigodot}$) red giant branch (RGB) stars \citep[see a review in][]{Serenelli_2017}. Initially, these stars ascend the red giant branch, increasing in brightness, as hydrogen is fused in a thin shell surrounding an inert helium core sustained by electron degeneracy pressure. Upon reaching a critical temperature, helium fusion starts, leading to a sudden temperature rise. This event, known as a helium flash, decreases electron degeneracy in the core, causing it to expand and produce a sudden increase in luminosity before the star's brightness fades, and it descends to the horizontal giant branch. This flash occurs at a characteristic magnitude and allows these stars to be used as a standard candle. Importantly, the TRGB luminosity is largely insensitive to metallicity and stellar mass in the Cousins $I$-band for metal-poor stars (\citealp{Da_Costa_1990, Lee_1993, Serenelli_2017}). The slope as a function of colour becomes more significant in other bands, and also with increasing metallicity, where the tip decreases in brightness and gets redder.

There have been a number of attempts to obtain a zero-point calibration of the TRGB luminosity and measure distances in the local universe (e.g \citealp{Lee_1993, Madore_2009, Hatt_2017, Freedman_2019}). Initially, the TRGB was observed using colour-magnitude diagrams of halo stellar populations in nearby galaxies such as M31 and M33 \citep{Mould_1986}. A new approach was then developed which involved the implementation of Sobel edge detection, a first-derivative, digital filter (\citealp{Lee_1993, Sakai_1997}), with a zero-sum kernel [${-2, 0, 2}$] convolved with a luminosity function (LF). This determines any discontinuity which corresponds to a sharp change in the LF at the location of the TRGB luminosity ($M^{TRGB}_{I}$). Different approaches to binning, Sobel kernels, and smoothing the LF have been explored \citep[see a review in][]{beaton2018oldaged}.

Direct measurement of $M_{I}^{TRGB}$ has only recently become possible, with the availability of milliarcsec parallax measurements from Gaia \citep{Gaia_2022}. Previously, the calibration of the TRGB was reliant on other calibrators, such as the Small Magellanic Cloud (SMC), the Large Magellanic Cloud (LMC), NGC 4258, and Milky Way globular clusters. \citet{Soltis_2021} measured the trigonometric parallax of $\omega$ Centauri using Gaia Early Data Release (EDR3), finding a TRGB magnitude, $M_{I}^{TRGB}$ = $-3.97 \pm 0.06$ mag. \cite{Mould_2019} utilised Gaia DR2 parallaxes and SkyMapper photometry of Milky way red giants to obtain a geometric calibration of the tip luminosity $\sim -4 $ mag in Cousins $I$. Additionally, \cite{Li_2022} used Gaia EDR3 to model the Milky Way stellar LF using a maximum likelihood approach and determined $M_{I}^{TRGB} = -3.91 \pm 0.05$ (stat) $\pm 0.09$ (sys) mag. 

While Gaia offers a direct measurement of the TRGB, it is still subject to factors that change with distance and the galaxy's environment. The key astrophysical elements that limit precision are extinction resulting from dust, crowding, and metallicity effects as discussed in \cite{Freedman_2021}.
To minimize these sources of error, it is important to focus on the halo regions of a galaxy where crowding is less of an issue, the impact of dust is minimal and can be corrected for, and limiting the colour range to include only bluer, metal-poor stars in the $I$-band to help mitigate metallicity effects on the slope of the TRGB.

Our work aims to obtain a geometric calibration of the TRGB using Gaia DR3 parallaxes. In Section \ref{Section2}, we first discuss the importance of Gaia astrometry and the necessary parallax corrections. Then we describe each of the photometric surveys used in our analysis. In Section \ref{Section3}, we first define our sample of Milky Way halo stars. Then we transform the photometry in our surveys to the Cousins photometric system and construct a smoothed luminosity function before applying Sobel edge detection to measure the TRGB luminosity. We show how this approach introduces a systematic offset, and how this bias can be corrected for by using simulations with PARSEC isochrones, to obtain measurements of $M_{I}^{TRGB}$. In Section \ref{Section4}, we analyse the main results, and compare with other TRGB zero-point calibrations, before discussing limitations, future work, and the impact on cosmology.

\begin{figure*}
    \centering  \includegraphics[width=\textwidth]{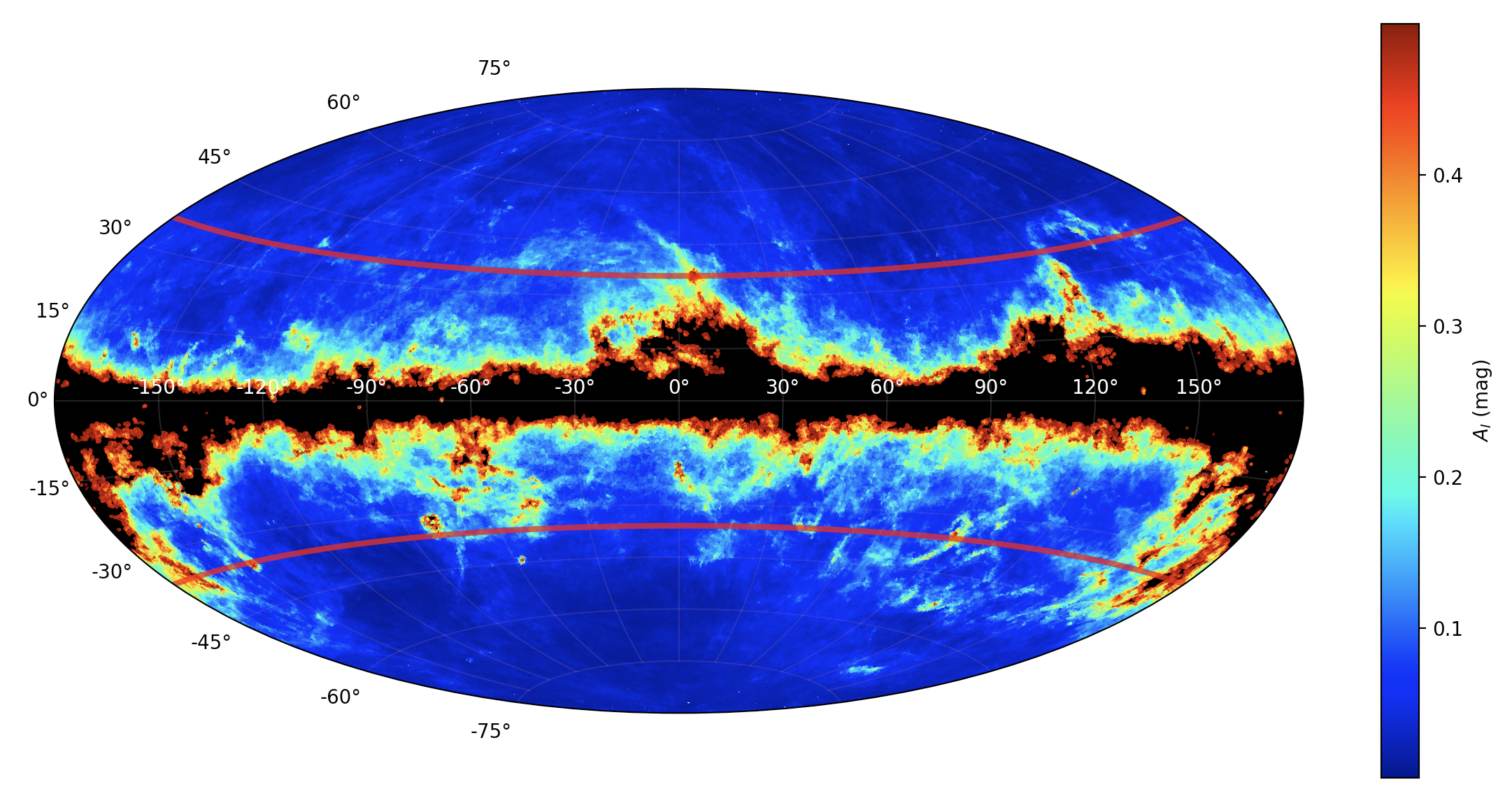}
    \caption{Galactic map of our Gaia DR3 sample, containing 7,733,106 stars. Higher values of reddening due to dust are evident around the thick disk and begin to drop off with increasing Galactic latitude. The red lines show our selection cut, where a Galactic latitude cut of $|b| > 36$, reduces the sample to 658,011 stars. This plot justifies our more conservative cut to focus on Milky Way halo stars, which tend to be more metal-poor and less impacted by dust and crowding. A more specific cut, where $A_{I} < 0.10$ mag is discussed in Section \ref{Section4}.}
    \label{fig:dust}
\end{figure*}

\section{Surveys}\label{Section2}
In this Section we describe the surveys, both ground and space-based, that we have utilised in this study to obtain astrometry and photometry.

\subsection{Astrometry}

The European Space Agency's (ESA) Gaia mission is obtaining a three-dimensional map of the Milky Way by measuring the parallaxes and proper motions for over 1.8 billion sources \citep{Lindegren_2020}.
On June 13th, 2022, Gaia Data Release 3 (Gaia DR3) was released \citep{Gaia_2022}. Compared to Gaia DR2, DR3 offers an increased number of parallaxes, and more precise measurements of parallax by a factor $\sim$ 0.8 \citep{Lindegren_2020}. 
Encompassing the range of the TRGB ($9 < G < 14$), the median parallax uncertainties are between 0.02-0.03 mas \citep{Lindegren_2020}. This enables a precise geometric calibration of the TRGB, which propagates directly to a systematic uncertainty in the value of $H_{0}$.

The precision of Gaia astrometry is subject to various instrumental effects that induce a systematic offset in observed parallaxes. 
This was evident in Gaia DR2, where a parallax zero-point offset correction was required \citep{Lindegren_2018}. For Gaia DR3, we follow the suggested zero-point parallax corrections described by \cite{Lindegren_2020} which are detailed position, apparent magnitude, and colour-dependent set of corrections. However, these corrections were obtained using 1.6 billion quasars which are much fainter ($G$ $>$ 19) than our RGB stars. Importantly, it was uncovered that there exists a systematic offset after the Lindegren corrections and has been probed by various studies (\citealp{Riess_2021, Zinn_2021, Huang_2021, Ren_2021, Flynn_2021, Li_2022}). In our region of interest ($10 < G < 13$ mag), this offset is negative and the  parallaxes are underestimated \citep{El-Badry_2021}. For our analysis, we use the zero-point correction obtained by \cite{Flynn_2021}, who find after the initial Lindegren corrections, a median parallax offset $\approx$ 10 $\mathrm{\mu}$as for bright stars (G $<$ 11). This was uncovered using the parallaxes of stars in open and globular clusters and is in agreement with other studies. 

In summary, the corrected parallax ($\varpi_{c}$) is obtained from the measured parallax ($\varpi$), $Z$ represents the Lindgren parallax bias correction, and we subtract the additional offset correction of 10 $\mathrm{\mu}$as.

\begin{equation}
    \varpi_{c} = \varpi - Z - 10~\mathrm{\mu} as
\end{equation}

\subsection{Photometry}
\subsubsection{Gaia}

Gaia DR3 now contains low-resolution spectrophotometry with flux calibration for $\approx$ 220 million sources, between 330--1050 nm in a range of filters \citep{Gaia_2022_specphot}. The synthetic photometry is then transformed and standardised to different photometric systems which include Johnson-Cousins ($UBVRI$), by using a collection of $\sim$ 100 million  secondary standards.

\subsubsection{SkyMapper}

The SkyMapper Southern Survey was conducted at the 1.35m SkyMapper Telescope at Siding Spring Observatory in Australia \citep{Onken_2019}. It provides astronomical data from $-90$ to +12 deg declination, covering most of the southern sky ($\approx$ 24,000 square degrees). Observing was undertaken between March 2014 and October 2019 across six optical filters: $uvgriz$ \citep{Bessell_2011}. The sources range from 8 to 22 AB magnitudes and saturation effects begin to arise for brighter objects, where $i$ $<$ 9 mag.

\subsubsection{APASS}
The American Association of Variable Star Observers (AAVSO)  Photometric All-Sky Survey began in 2010, at
the Cerro Tololo Inter-American Observatory (CTIO) in Chile and the Dark Ridge Observatory in New Mexico \citep{Henden_2016}. APASS Data Release 9 (DR9) contains approximately 62 million objects covering 99$\%$ of the sky. The filters include Sloan $g'$, $r'$, $i'$, B and V. The photometry between $7 < m < 17$ is precise and reliable, covering an important magnitude range and can be directly compared with similar photometric surveys. We utilise APASS DR9 and not DR10, which has some issues involving all-sky coverage, and larger photometric errors. The saturation point for APASS DR9 is $\sim V = 10$ mag.

\subsubsection{Pan-STARRS}

The Panoramic Survey Telescope and Rapid Response System (Pan-STARRS), surveyed the sky north of declination $-30$ degrees in five broad-band filters $(g,r,i,z,y)$ between 2010 and 2015 (\citealp{PanSTARRS_2016,Flewelling_2020}). The Pan-STARRS (PS1) survey utilised the 1.8m telescope and 1.4 Gigapixel camera situated in Haleakala, Hawaii.

\subsubsection{ATLAS Refcat2}
The Asteroid Terrestrial-impact Last Alert System (ATLAS) \citep{2018Tonry} provides an all-sky reference catalogue, containing around one billion sources as faint as an apparent magnitude, $m \sim$ 19 mag. The ATLAS Refcat2 catalogue is a combination of various photometric survey's which include Pan-STARRS, APASS, SkyMapper, Tycho-2, and the Yale Bright Star Catalogue. The photometry for each object is then transformed onto the same photometric system (Pan-STARRS: $g,r,i,z,y$).

\section{Methodology}\label{Section3}

\subsection{Sample Selection}
To obtain our initial sample of Milky Way RGB stars we use the below query, which contains 7,733,106 stars from Gaia DR3\footnote{Data retrieved at https://gea.esac.esa.int/archive}, and can be accessed online. \\

\noindent \textit{SELECT * FROM $gaiadr3.gaia\_source$ \\
WHERE $(b >= 36$ OR $b <= -36)$ \\
AND $phot\_g\_mean\_mag <= 14$ \\
AND $(1 <= bp\_rp < 3)$ \\
AND $parallax\_over\_error > 5$ \\
AND $parallax > 0$  \\
AND $ruwe < 1.4$} \\
%AND $astrometric\_params\_solved > 5$ \\

We first select high latitude stars ($|b| > 36$), limiting the impact of dust \citep{Freedman_2020}. We now have a high latitude colour-magnitude diagram (CMD) with TRGB stars in the halo regions, which helps limit the contamination of the sample by metal-rich disk stars. A colour and $G$-band cut is added to focus on selecting RGB stars in our region of interest. We define the fractional parallax error (FPE) as $\sigma_{\varpi}/\varpi$. Objects with a FPE $>$ 0.2 are removed due to the linear propagation of these errors into magnitudes becoming significant.
Our query also follows the suggestion in \cite{Lindegren_2020}, to use the goodness-of-fit parameter ``Renormalized Unit Weight Error'' (ruwe) $< 1.4$ to filter out stars with poor astrometric solutions. This value is determined as a safe cut for obtaining high-quality parallaxes.

%\begin{table*}
%\centering
%\caption{Example stars for each of the photometric surveys, containing sky position, %corrected parallax, apparent magnitude, absolute magnitude, colour...}
%\begin{tabular}{c|c|c|c|c|c|c|c|c}
%\hline
%Photometric Survey & Gaia ID & RA (deg) & DEC (deg) & $BP-RP$ (mag) & %$\overline{\omega}_{corrected}$ (mas) & $\sigma$ (mag) & I (mag) & $M_{I}$ (mag)   \\
%\hline
%SkyMapper & 1011884485 & 43.578 & -39.7781 & 1.4 & 0.03 & 0.003 & 11.4 & -4.01 \\
%\hline
%\end{tabular}
%\label{Table:example_stars}
%\end{table*}

\subsection{Photometric transformation to Cousins $I$}\label{transformations}

We then cross-match our sample using the Gaia Archive and TOPCAT \citep{TOPCAT_2011}, with a matching radius of 1'' for each of the photometric catalogues; SkyMapper DR3, APASS DR9 and ATLAS Refcat2. Gaia DR3 synthetic photometry is obtained by directly matching with the Gaia Synthetic Photometry Catalogue (GSPC) using each Gaia `source$\_$id'.

To obtain $M_{I}$ for each star, we need to convert each survey respectively to the Cousins $I$ photometric system, where the TRGB is the flattest. To account for host galaxy dust, we apply extinction corrections obtained from a 2D map of dust reddening \citep{Schlegel_1998}, and more recently updated by \cite{Schlafy_2011}. We can then convert to extinction in the 
$I$ band using known extinction laws \citep{Cardelli_1989} to obtain $A_{I}$. We use the zero point offset between \cite{Schlafy_2011} and \citep{Planck_2014} for objects in our sample, to adopt a conservative systematic error of 0.01 mag in $A_{I}$. The Galactic latitude and longitude of stars in our initial sample are shown in Figure \ref{fig:dust}.

We derive colour transformations using a collection of Stetson standard stars cross-matched with Gaia DR3 \citep{Pancino_2022}. For each survey, we plot the difference between $i$ and Cousins $I$ (Figures \ref{fig:atlas_offset}, \ref{fig:apass_offset}, \ref{fig:sky_offset}), and provide a linear fit within our region of interest for correcting each object.

\begin{equation}
    I = i_{SM} - 0.0562\,(BP-RP) - 0.3727; \\ \sigma = \pm0.0159
\end{equation}

\begin{equation}
    I = i_{AP} - 0.0756\,(BP-RP) - 0.3534;\\ \sigma = \pm0.0294
\end{equation}

\begin{equation}
    I = i_{AT}- 0.1018\,(BP-RP) - 0.3564;\\ \sigma = \pm0.0200
\end{equation}

%\begin{equation}
%    I = i_{SM} - 0.45 - A_{i}
%\end{equation}
%\begin{equation}
%    I = i_{PS} - 0.367 - 0.149(g_{PS} - r_{PS}) - A_{i}
%\end{equation}

\begin{equation}
    M_{I} = I + 5\mathrm{log(\varpi_{c})} - 10 - A_{I}
\end{equation}

where $M_{I}$ is the absolute magnitude on the Cousins system, $I$ is the dereddened Cousins apparent magnitude, $A_{I}$ is the amount of extinction in the $I$-band and $\varpi_{c}$ is the corrected parallax in mas. Table \ref{Table:systematics} shows the systematic uncertainties for each transformation, which have been calculated by determining the standard deviation of the overall offset for each sample.
We note that the Gaia synthetic photometry has already been transformed to Cousins $I$. Using Table 2 and Figure 9 in \cite{Gaia_2022_specphot}, we add the photometric uncertainties and calibration dispersion ($\pm$0.0137 mag) in quadrature and adopt a 
systematic uncertainty 
of $\pm$0.005 mag. 

We then make a selection cut to focus on stars within $-5 < M_{I} < -2$. We also make a cut where $A_{I} < 0.1$ mag to omit any stars significantly affected by extinction. A further cut on $I$ is implemented to avoid incompleteness in our samples ($8.6 < I < 11.7$) and remove over-saturated stars with known large photometric uncertainties ($\sigma_{i} < 0.1$ mag). Further survey specific photometric quality cuts include SkyMapper: $i_{nimaflags}$ < 15 and Gaia: $c_{\star}$ < 0.05 \citep{Gaia_2022_specphot}.

\begin{figure*}
    \centering
    \includegraphics[width=\textwidth]{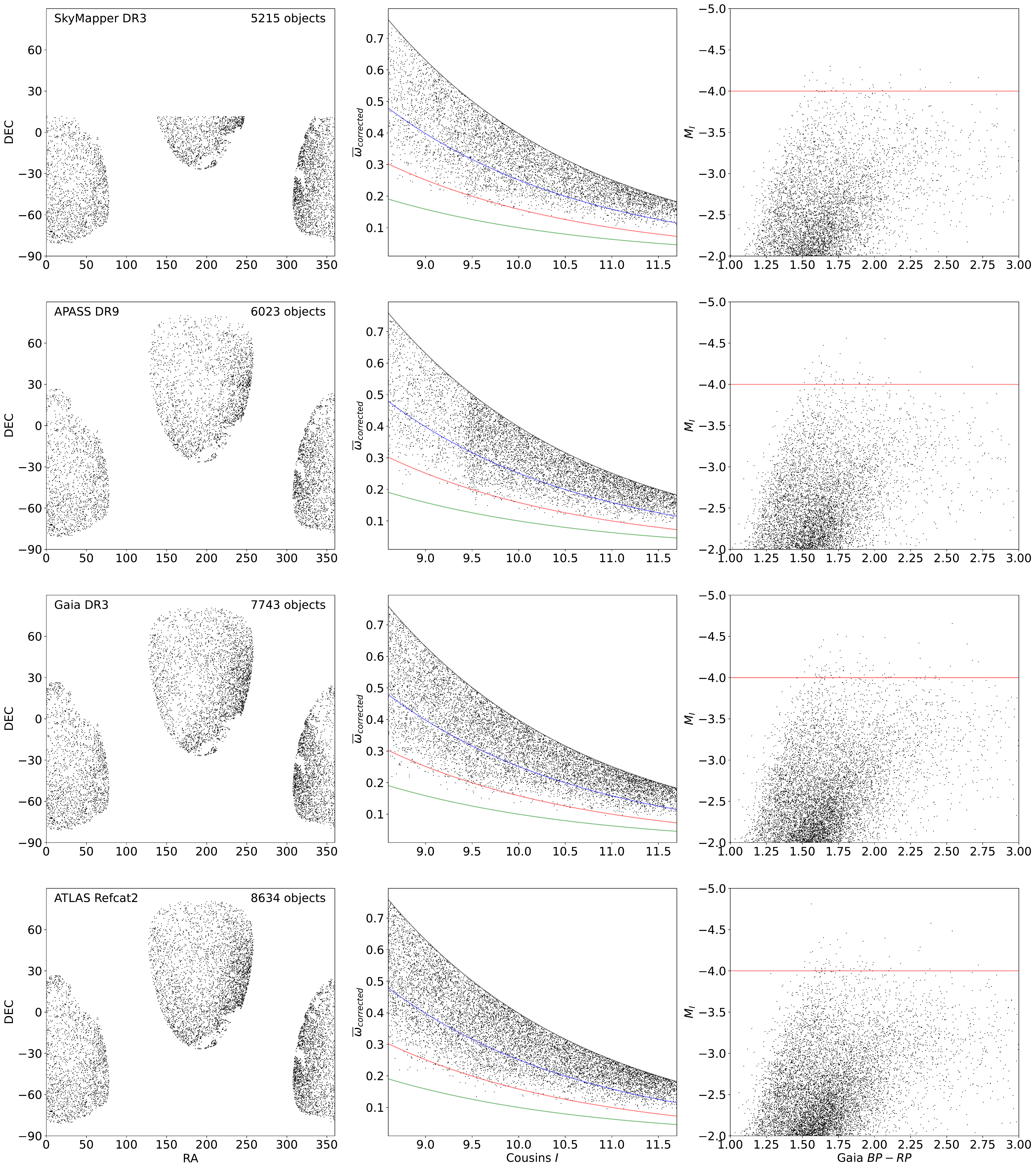}
    \caption{Survey comparison with a magnitude cut ($-5 < M_{I} < -2)$, fractional parallax cut (FPC) $<$ 0.1, $A_{I} <$ 0.1 and $\sigma_{i} <$ 0.1. \textbf{Left}: Sky position for each survey, where SkyMapper covers the Southern Hemisphere and APASS/ATLAS/Gaia are all-sky surveys. \textbf{Centre}: Corrected parallaxes plotted against dereddened Cousins $I$, where the black line corresponds to $M_{I} = -2$ mag, the green line $M_{I} = -5$ mag, while the red line is an approximate location of the TRGB at $M_{I} \approx -4$ mag. \textbf{Right}: CMD with the red line showing the approximate location of the TRGB.}
    \label{fig:surveys}
\end{figure*}

In Figure \ref{fig:surveys}, we plot the distribution of stars on the sky,  dereddened Cousins $I$ against the corrected parallax, and a CMD for each of the samples, where the drop-off in the density of stars can be seen around $M_{I} \approx-4 $ and represents the approximate location of the TRGB. \cite{Madore_2009} suggest that between 400 and 500 stars within 1 mag of the TRGB is needed for an accurate measurement of the TRGB to within 0.1 mag. This sets the threshold for the minimum number of objects needed in our sample between $-5 < M_{I} < -3$. The parallaxes fall between 0.1 and 0.75 mas, indicating distances within the range of 1.3 to 10.5 kpc. These values reinforce our original selection criteria of selecting halo stars, as the median distance of the sample $\approx$ 3.3 kpc.

\subsection{Sobel Edge Detection}

The Sobel filter has been implemented across a range of studies in detecting the TRGB peak luminosity. In principle, you can bin by $M_{I}$ and construct a histogram. A $[-1, 0, 1]$ kernel is used for convolution, and the Sobel response function is weighted to reduce Poisson noise \citep{Mager_2008}. But, this is sensitive to how the bins are defined.

The better approach is to smooth the LF using `Gaussian-windowed, Locally Weighted Scatter Smoothing’ (GLOESS) \citep{Persson_2004}. While there are limitations with this approach, it is less sensitive to binning choices and has the advantage of increasing the S/N of the distribution to obtain a more precise result. We instead use a kernel-density estimator (KDE) to smooth our LF. While similar to GLOESS, our approach removes the bin size variable and the most important parameter is the smoothing kernel.

Importantly, while a larger smoothing kernel can improve precision, it introduces a more significant systematic error (see Figure \ref{fig:sim_offsets}), and has also been uncovered in recent TRGB studies.  This bias is driven by the shape of the LF, the photometric scatter and choice of smoothing kernel. We discuss our methodology in Section \ref{simulations}, to calibrate out and remove this systematic effect.

\begin{table}
\centering
\caption{Our analysis accounts for a number of systematic factors, including photometric transformation to Cousins $I$, extinction ($A_{I}$), isochrone metallicity, AGB contamination, and the smoothing kernel used in our simulations. We estimate the systematic total to be 0.031 mag, where we add the average of the four photometric transformations ($I_{APASS}$, $I_{Gaia}$, $I_{ATLAS}$, $I_{SkyMapper}$) and them combine the other errors in quadrature.}
\begin{tabular}{c|c}
\hline
Source & Systematic error (mag)   \\ 
\hline
Parallax offset correction &  $\pm$0.004 
($\pm$2$\mathrm{\mu}$as) \\
$A_{I}$ & $\pm$0.01 \\ 
$I_{APASS}$ & $\pm$0.023 \\
$I_{Gaia}$ & $\pm$0.005 \\
$I_{ATLAS}$ & $\pm$0.030 \\ 
$I_{SkyMapper}$ & $\pm$0.016 \\
Smoothing kernel (sim) & $\pm$0.02  \\ 
Isochrone (sim) & $\pm$0.01 \\ 
AGB contamination (sim) & $\pm$0.005  \\ 
\hline
Estimated systematic contribution & $\pm$0.031\\
\hline
\end{tabular}
\label{Table:systematics}
\end{table}

\begin{table*} 
\centering 
\caption{TRGB geometric calibration using each of the photometric surveys. A FPC of 0.10, was implemented for each survey with a smoothing kernel of 0.09 mag. We also show the number of stars in each sample, median $\sigma_{M_{I}}$ and offset correction.}
\begin{tabular}{c|c|c|c|c|c}
\hline
Photometric Survey & $\#$ Stars & $\sigma_{M_{I}}$ & Offset Correction & $M_{I}^{TRGB}$ &$\sigma$\\ 
\hline
SkyMapper DR3 & 5215 & 0.130 & 0.087 & $-4.007$ & $\pm0.034$ (stat) $\pm0.030$ (sys) \\ 
APASS DR9 & 6023 & 0.140 & 0.093 & $-4.032$ & $\pm0.036$ (stat) $\pm0.034$ (sys) \\
Gaia DR3 & 7743 & 0.121 & 0.083 & $-4.035$ & $\pm0.037$ (stat) $\pm0.026$ (sys) \\
ATLAS Refcat2 & 8634 & 0.124 & 0.084 & $-4.044$ & $\pm0.022$ (stat) $\pm0.040$ (sys) \\
\hline  
Full ensemble & 9148 & 0.121 & 0.082 & $-4.042$ & $\pm0.041$ (stat) $\pm0.031$ (sys)\\
\hline
\end{tabular}
\label{Table:trgb_calibration}
\end{table*}

\subsection{Simulations of TRGB detection with isochrones} \label{simulations}

To examine and constrain the systematic errors in our Sobel edge detection method, we generate stellar mock samples using PARSEC (Padova and Trieste Stellar Evolutionary Code) CMD 3.7 \citep{Padova_2012,Marigo_2017}, to model the TRGB peak dispersion in different stellar populations. We choose to use single stellar population isochrones as we are not directly modelling the Milky Way CMD. The Milky Way is comprised of stars of varying ages, metallicities and distances, but our observational data, which focuses on stars at high galactic latitudes, limits our TRGB sample to generally contain older, more metal-poor halo stars. 

\begin{figure}
    \centering
    \includegraphics[scale=0.35]{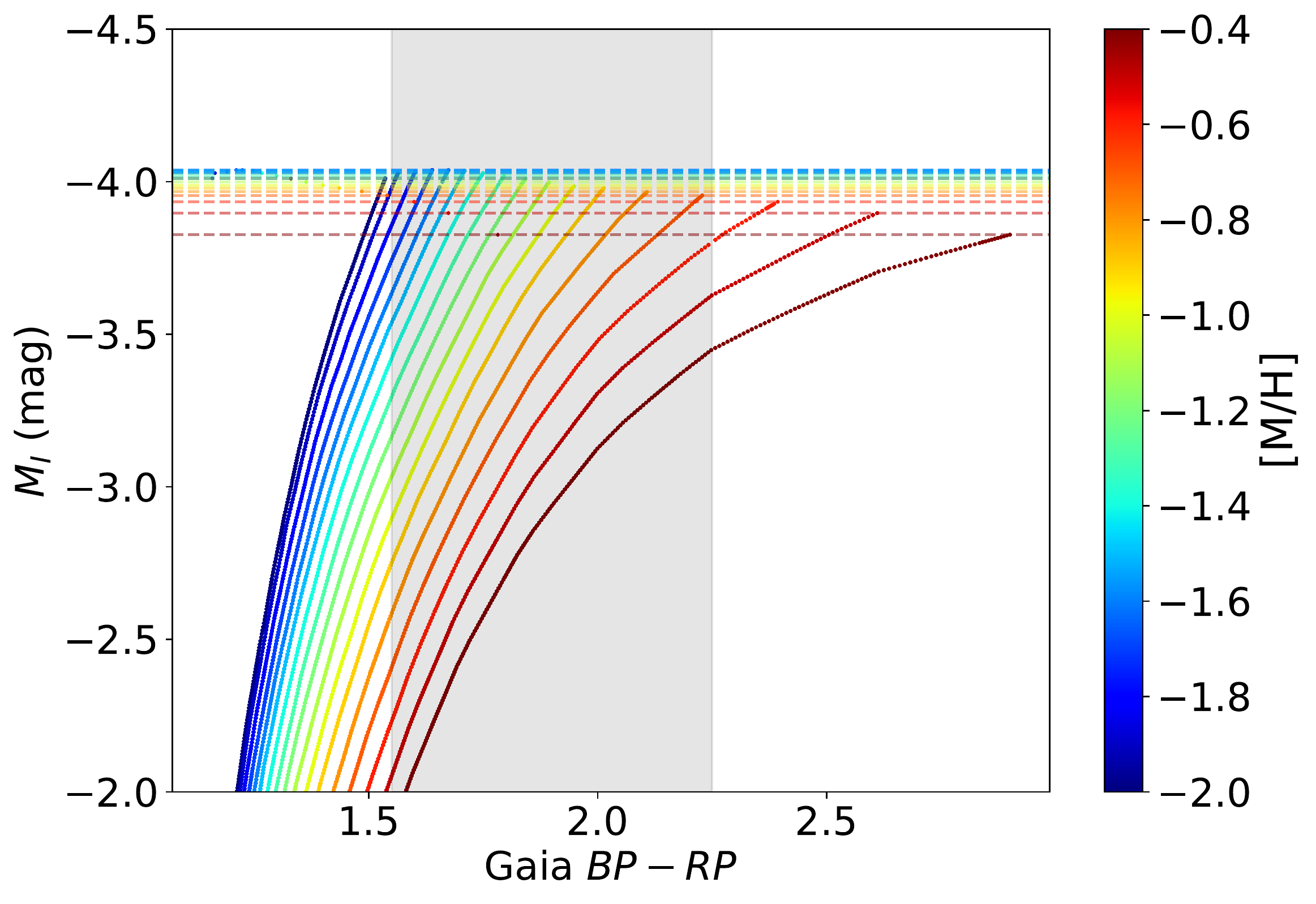}
    \caption{Isochrones with a  stellar age of 10 Gyr, and a range of metallicities 
    ($-2 < [M/H] < -0.4$). The slope of the TRGB is shown to be relatively flat within the colour range ($1.55 < BP-RP < 2.25$).}
    \label{fig:padova}
\end{figure}

The offset of Sobel edge detection applied to the smoothed LF is determined by the theoretical TRGB cutoff for each isochrone. Figure \ref{fig:padova} displays a range of RGB stellar tracks, showing that the TRGB slope remains relatively flat across the colour range of the isochrones. 

To define the photometric scatter for our simulated stars, we use the median photometric scatter ($\sigma_{M_{I}}$) for each observational sample. The sample size $N$, is constrained by the availability and precision of the parallaxes, as the errors linearly propagate through to magnitudes. Importantly, parallax errors shift the location of the edge by smoothing out the distribution, which means that we can control the degree of systematic error in our simulations. Quantitatively, if we apply a fractional parallax cut (FPC) of 0.25 this translates to $\sigma_{M_{I}}$ $\sim$ 0.54 mag, while a FPC of 0.1 translates to $\sigma_{M_{I}}$ $\sim$ 0.13 mag. While having a lower FPE provides advantages, it results in a smaller sample size, which can impact the accuracy of measuring the TRGB. We note that setting a lower FPC biases our sample toward brighter objects as larger uncertainties tend to arise for fainter objects which are omitted. We then model the relationship between $\sigma_{M_{I}}$ and $N$ for each of our samples for a given (FPC).

\begin{figure}
    \centering
    \includegraphics[scale=0.35]{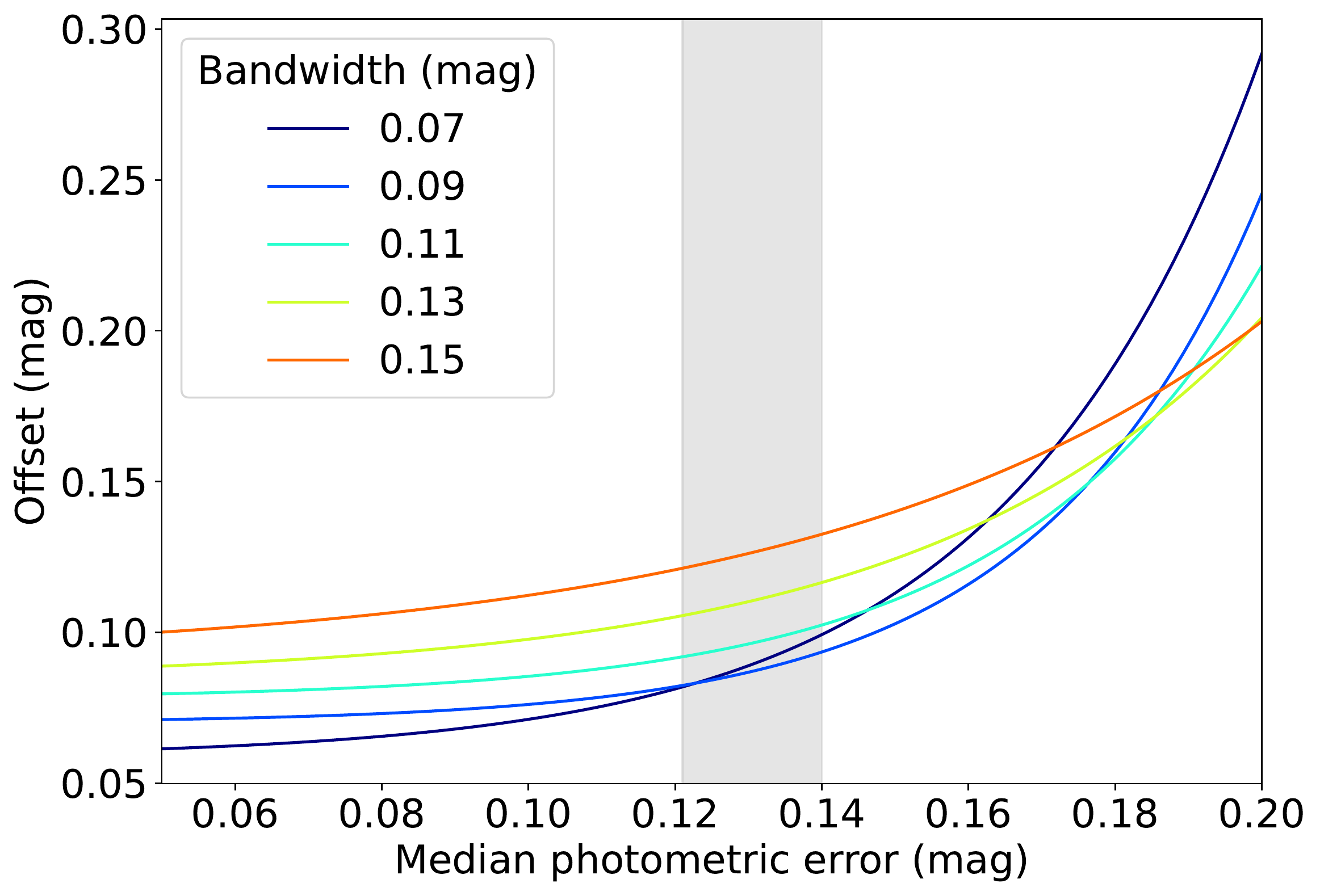}
    \caption{Simulations to obtain systematic offsets with increasing scatter, for a range of smoothing kernels (0.07$-$0.15 mag). The grey region corresponds to the range of median photometric scatter ($\sigma_{M_{I}}$) in Table \ref{Table:trgb_calibration}. Here we can justify our selection of using a smoothing kernel (0.09 mag), where the bias is minimal, given our typical photometric uncertainties. We can see that for low $\sigma_{M_{I}}$ the bias is driven by the scale of the smoothing kernel, while with increasing noise the bias becomes driven by $\sigma_{M_{I}}$ and larger smoothing kernels are needed to suppress the noise.}
    \label{fig:sim_offsets}
\end{figure}

We aim to estimate TRGB offsets as simply as possible by focusing on a single stellar population isochrone. However, we can obtain the spread in this systematic, depending on the choice of isochrone. We generate 10,000 stars to minimise Poisson systematics and examine the impact of isochrone metallicity on measuring the TRGB. This was repeated 500 times and we found that the scatter and offset of the simulated TRGB peak only slightly changed ($\sim$ 0.01 mag) with isochrone metallicity ($-2 < [M/H] < -0.4$).

We next choose an isochrone similar to our observational samples, with a stellar age of 10 Gyr and $[M/H]$ = $-1.6$ and generate stars between ${-5 < M_{I} < -2}$. \cite{Serenelli_2017} determined that there is little dependence on the age of TRGB stars older than 4 Gyr. We then convert Cousins ($V-I$) to Gaia ($BR-RP$), using the transformation obtained in Figure \ref{fig:colour}, with a sample of 717 standard stars \citep{Gaia_2022_specphot}.

The measurement of the TRGB tip in observational samples is affected by the presence of oxygen-rich asymptotic giant branch (AGB) stars, which lie not just above the TRGB but cross through it. Thus we need to quantify this impact in our simulations. To model this contamination, we add stars that are linearly generated between $-5 < M_{I} < -2$ and evaluate the impact of varying the contamination levels (2.5$\%$, 5$\%$, 10$\%$ of $N$). 
We find no significant difference in the scatter or offset of the TRGB magnitude and choose to conservatively add 5$\%$ of $N$ linearly distributed stars in each simulation to model AGB contamination. Similarly, \cite{Hatt_2017} found no significant impact on the detection of the TRGB discontinuity when varying the AGB/RGB ratio.

Another systematic error in our measurements is the width of the smoothing kernel, which offsets the TRGB peak. It is ideal to have a low $\sigma_{M_{I}}$, where the smoothing scale can be minimised. However, a larger smoothing kernel is needed to suppress increasing $\sigma_{M_{I}}$, and this relationship is shown in Figure \ref{fig:sim_offsets}.
For each of the simulated survey samples, we examined various smoothing kernel widths ranging from 0.07 to 0.15 mag. For smaller values of $\sigma_{M_{I}}$, the systematic offset is driven by the smoothing kernel. As $\sigma_{M_{I}}$ becomes large, the photometric scatter becomes the dominant factor. The shaded grey region in Figure \ref{fig:sim_offsets}, represents the range of $\sigma_{M_{I}}$ for the observational samples. For a given FPC/$\sigma_{M_{I}}$ or smoothing kernel, we can find the optimal systematic offset and apply the correction to our data. Hence we justify our choice of a smoothing kernel of 0.09 mag, where the systematic offset is minimal, given the photometric scatter, for use in our analysis.

In summary, the main sources of systematic error in our simulations are the choice of isochrone, the smoothing kernel bandwidth and contamination from AGB stars, which was found to have a negligible impact on the offset and scatter of the TRGB. Despite each isochrone having a different theoretical TRGB cut-off, the offset remains consistent, allowing for correction in our data. Although it is difficult to account for all systematic factors in our analysis, Table \ref{Table:systematics} lists those that we have taken into consideration. We choose to add the systematic error contributions in quadrature but note that this is a limitation in our approach, as some of these factors may be correlated.

\section{Results and Discussion}\label{Section4}

\subsection{Measuring the magnitude of the TRGB}

We can now apply Sobel edge detection on the luminosity function for each survey and incorporate an offset correction to determine $M_{I}^{TRGB}$. To estimate the statistical uncertainty, we bootstrap objects within 0.2 mag and measure the variation in the TRGB peak. This is repeated 10,000 times to determine an average $M_{I}^{TRGB}$. From our simulations, we utilise a smoothing kernel of 0.09 mag and FPC of 0.10 for each of the surveys. 

\begin{figure*}
    \centering
    \includegraphics[width=\textwidth]{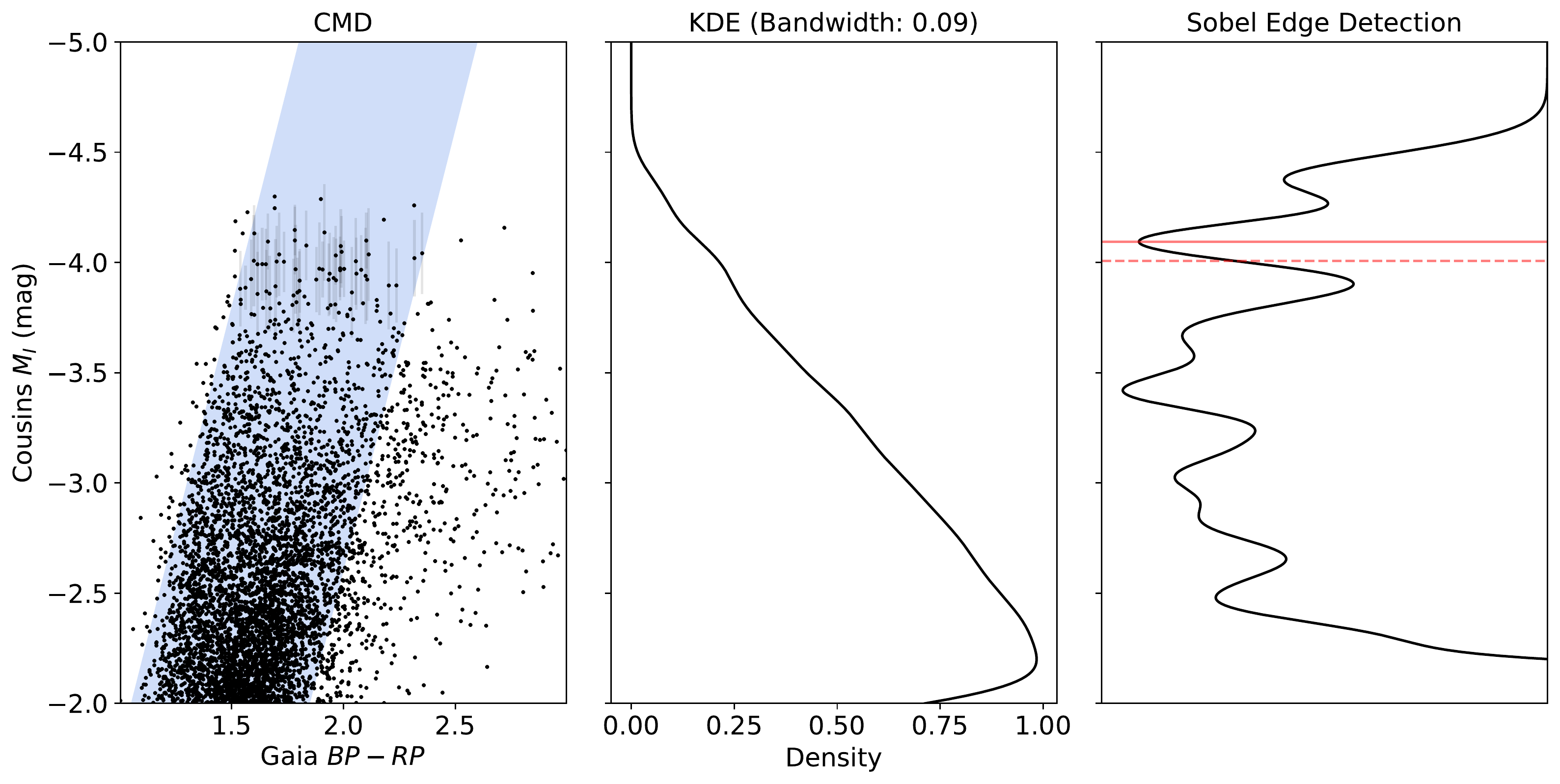}
    \caption{SkyMapper sample (5215 stars): \textbf{Left}: $M_{I}$ plotted against Gaia colour ($BP-RP$). The blue shaded box is the region we focus on, where the blue TRGB, defined as $1.55 < BP-RP < 2.35$, is relatively flat and limits the contamination of redder, more metal-rich stars. Errorbars are shown for stars around the TRGB. 
    \textbf{Centre}: Luminosity function derived using a kernel density estimator, with a smoothing kernel of 0.09 mag.
    \textbf{Right}: Response after applying Sobel edge detection, where the solid red line represents the detected peak and the dashed line is after the offset correction (0.087 mag) derived from simulations is applied. We find $M_{I}^{TRGB} = -4.007$ mag.}
    \label{fig:SKY}
\end{figure*}

\begin{figure*}
    \centering
    \includegraphics[width=\textwidth]{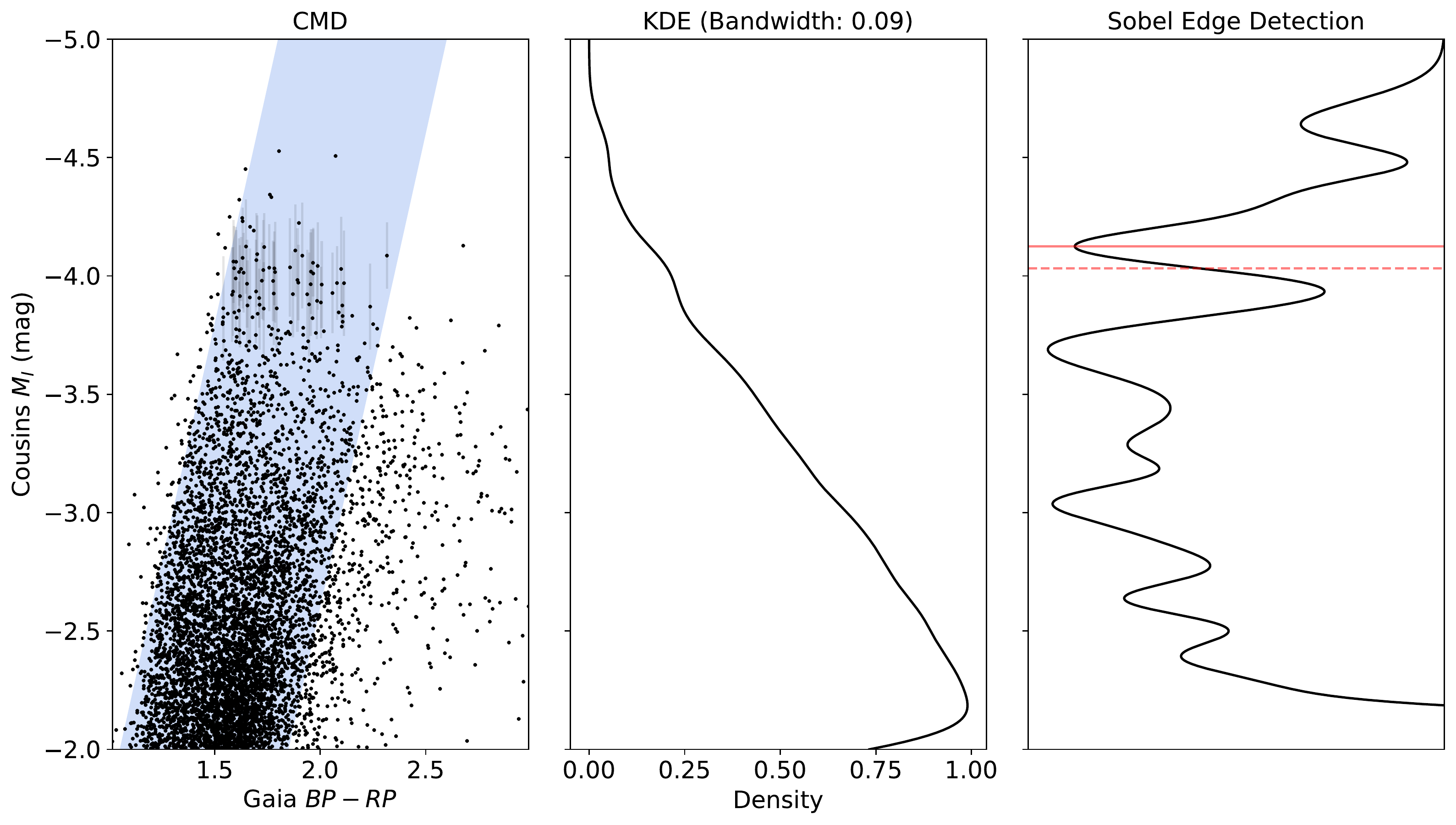}
    \caption{Same as Figure \ref{fig:SKY} but for the APASS sample, containing 6023 stars and an offset correction of 0.093 mag. We find $M_{I}^{TRGB} = -4.032$ mag.}
    \label{fig:APASS}
\end{figure*}

\begin{figure*}
    \centering
    \includegraphics[width=\textwidth]{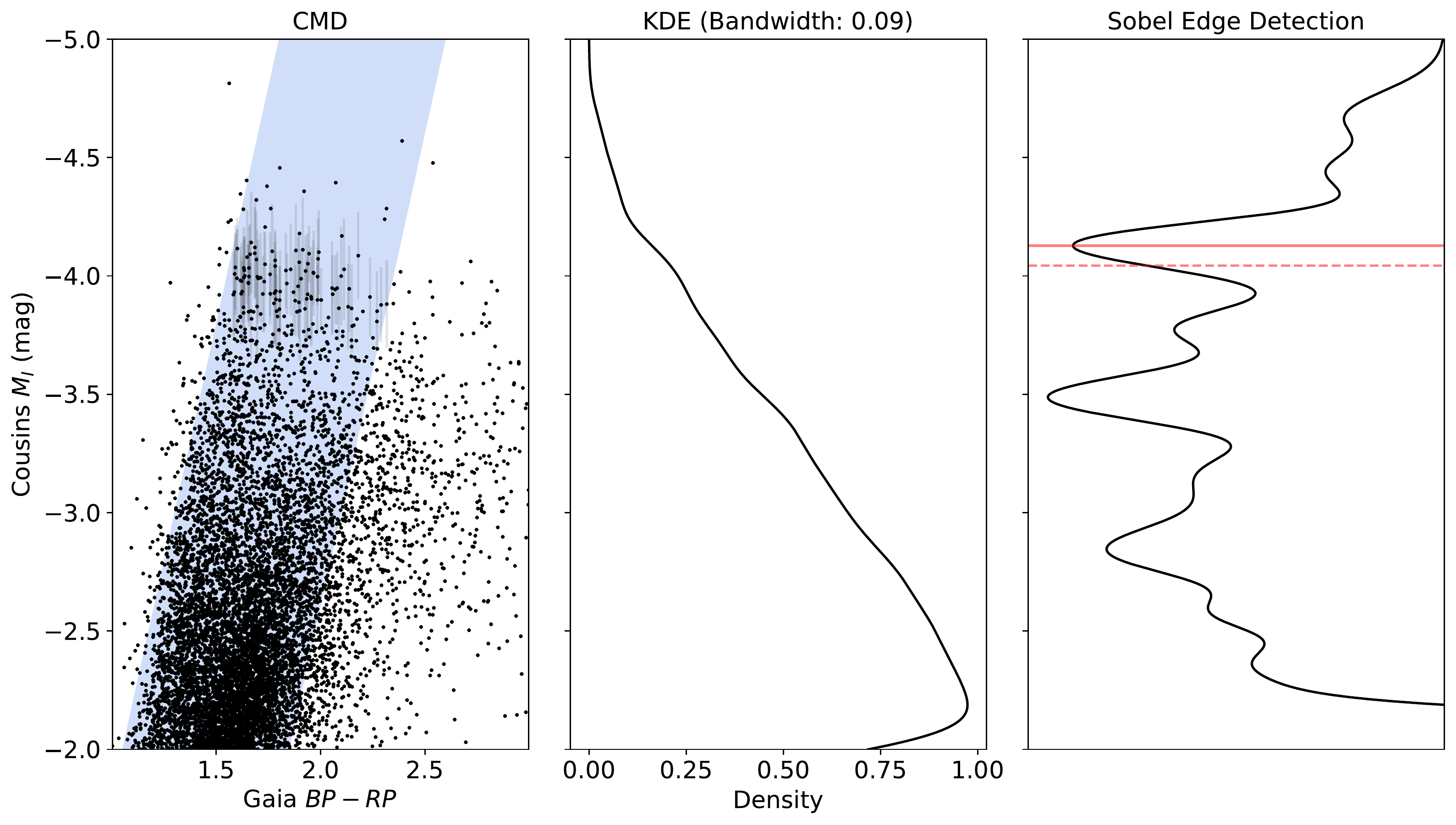}
    \caption{Same as Figure \ref{fig:SKY} but for the ATLAS sample, containing 8634 stars and an offset correction of 0.084 mag. We find $M_{I}^{TRGB} = -4.044$ mag.}
    \label{fig:ATLAS}
\end{figure*}

\begin{figure*}
    \centering
    \includegraphics[width=\textwidth]{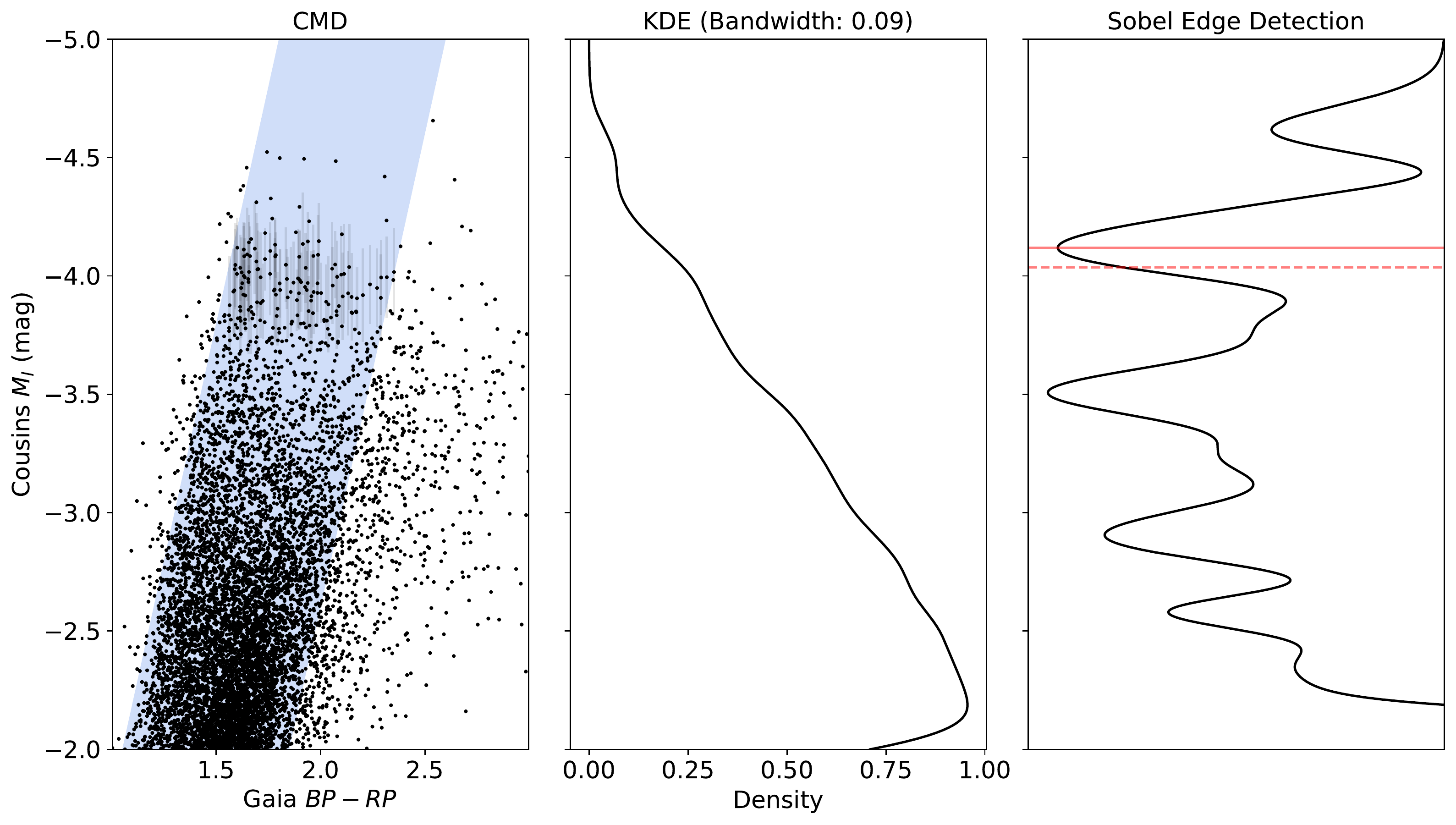}
    \caption{Same as Figure \ref{fig:SKY} but for the Gaia sample, containing 7743 stars and an offset correction of 0.083 mag. We find $M_{I}^{TRGB} = -4.035$ mag.}
    \label{fig:GAIA}
\end{figure*}

We adopt a similar approach to other TRGB studies, where we define a region known as the blue TRGB, where $M_{I}^{TRGB}$ is relatively flat and varies little with colour. We want to focus on selecting bright metal-poor RGB stars. \cite{Hoyt_2021} found a flat $I$-band TRGB magnitude between $1.45 < (V-I) < 1.95$ mag. Similarly, \cite{Freedman_2020} defined a colour range $1.4 < (V-I) < 2.2$ mag, corresponding to $-1.4 < [M/H]< -0.6$. We select the colour range $1.55 < BP-RP < 2.25$, with a slope $-$4 mag mag$^{-1}$, defined by the blue shaded region in Figures \ref{fig:SKY}, \ref{fig:APASS}, \ref{fig:ATLAS}, \ref{fig:GAIA}. This region fits the metal-poor isochrones shown in Figure \ref{fig:padova}. Without this selection, the TRGB peak is less well defined, with the introduction of redder, more metal-rich stars which induce a more significant slope \citep{Hoyt_2021}, i.e. the TRGB luminosity becomes fainter with redder colour. As in previous work, it is important to avoid the inclusion of such stars in our sample. In our analysis, we do not observe a single well-defined peak across our CMDs when using Sobel edge detection. This can be attributed to observational scatter which is accounted for in our simulations. However, within the region of the TRGB we can distinguish $M_{I}^{TRGB}$ and its scatter, as the other peaks are driven by noise in the CMD and are expected to be at fainter magnitudes than $M_{I}^{TRGB}$.

The TRGB luminosity has been calibrated using a variety of different approaches. Some of the more recent calibrations are summarised in Table \ref{Table:trgb_surveys}, which includes TRGB measurements of the SMC, LMC, the spiral galaxy NGC 4258 and $\omega$ Centauri. We note that the calibrations rely on distances derived using different anchors, including masers, parallaxes and double-eclipsing binary stars. \cite{Freedman_2021} obtain a weighted average zero-point calibration of $-4.042$ mag using four independent calibrators (SMC, LMC, $\omega$ Cen, NGC 4258).

\begin{table*}
\centering
\caption{Recent TRGB luminosity calibrations along with $H_{0}$ measurements where available. Each study was converted from F814W to I if applicable by adding 0.0068 mag \citep{Freedman_2019}. The sources of each zero-point calibration include Gaia parallaxes, masers and detached eclipsing binaries (DEBs). We note that the parallax measurements in Gaia EDR3 are the same as Gaia DR3. NGC 4258 is a mega-maser host galaxy and DEBs are used to accurately measure distances in the LMC to 1$\%$ \citep{Pietrzynski2013}.}
\begin{tabular}{c|c|c|c|c|c|c}
\hline
& Calibrator & $M_{I}$ & $\sigma$ & $H_{0}$ & $\sigma$ \\ 
 \hline
 \cite{Hoyt_2023} & LMC & $-4.038$ & $\pm$0.012 (stat) $\pm$0.032 (sys) \\
\cite{Hoyt_2023} & SMC & $-4.050$ & $\pm$0.030 (stat) $\pm$0.039 (sys) \\
 \cite{Anderson_2023} & LMC & $-3.979$ & $\pm$0.011 (stat) $\pm$0.028 (syst) & 71.8 & $\pm$1.5\\
\cite{Li_2022} & MW (Gaia EDR3) & $-3.91$ & $\pm$0.05 (stat) $\pm$0.09 (sys)\\
\cite{Anand_2022} & NGC 4258 & $-$4.00 & $\pm$0.04 & 71.5 & $\pm$1.8\\
%\cite{Dhawan_2022} & NGC7814 & & & 77.58 & $\pm$6.1\\
%\cite{Apellaniz_2021} & $\omega$ Cent & -4.042 & $\pm$0.015 (stat) $\pm$0.035 (sys) & 69.8 & $\pm$0.6 (stat) $\pm$1.6 (sys) \\
\cite{Freedman_2021} & (SMC, LMC, $\omega$ Cen, NGC 4258) & $-$4.042 & $\pm$0.015 (stat) $\pm$0.035 (sys) & 69.8 & $\pm$0.6 (stat) $\pm$1.6 (sys) \\
\cite{Jang_2021} & NGC 4258 & $-$4.043 &  $\pm$0.028 (stat) $\pm$0.048 (sys) \\
\cite{Soltis_2021} & $\omega$ Cen (Gaia EDR3) & $-$3.97 & $\pm$0.06 & 72.1 & $\pm$2.0 \\
\cite{Freedman_2020} & LMC & $-$4.047 & $\pm$0.022 (stat) $\pm$0.039 (sys) & 69.6 & $\pm$1.1 (stat) $\pm$1.7 (sys) \\
\cite{Cerny_2020} & $\omega$ Cen & $-$4.056 & $\pm$0.02 (stat) $\pm$0.10 (sys) \\
\cite{Górski_2020} & LMC & $-$3.956 & $\pm$0.04 \\
\cite{Capozzi_2020} & $\omega$ Cen (Gaia DR2) & $-$3.96 & $\pm$0.05 \\
\cite{Yuan_2019} & LMC & $-$3.963 & $\pm$0.046 & 72.4 & $\pm$2.0 \\
\cite{Mould_2019} & MW (Gaia DR2) & $-$4.00 & $\pm$0.076 \\
%\cite{Reid_2019} & NGC 4258 & -4.003 & $\pm$0.04 & 71.1 & $\pm$1.9\\
\cite{Jang_2017} & NGC 4258, LMC & $-$4.016 & $\pm$0.058 \\

\hline
This Study & MW (Gaia DR3) & $-$4.042 & $\pm$0.041 (stat) $\pm$0.031 (sys)\\
\hline
\end{tabular}
\label{Table:trgb_surveys}
\end{table*}

Our $M_{I}^{TRGB}$ values, offset corrections, the number of stars and median scatter are summarised in Table \ref{Table:trgb_calibration}. The $M_{I}^{TRGB}$ measurements range from $-4.007$ to $-4.044$ mag, where each has a similar offset correction derived from our isochrone simulations. 

\begin{figure*}
    \centering
    \includegraphics[width=\textwidth]{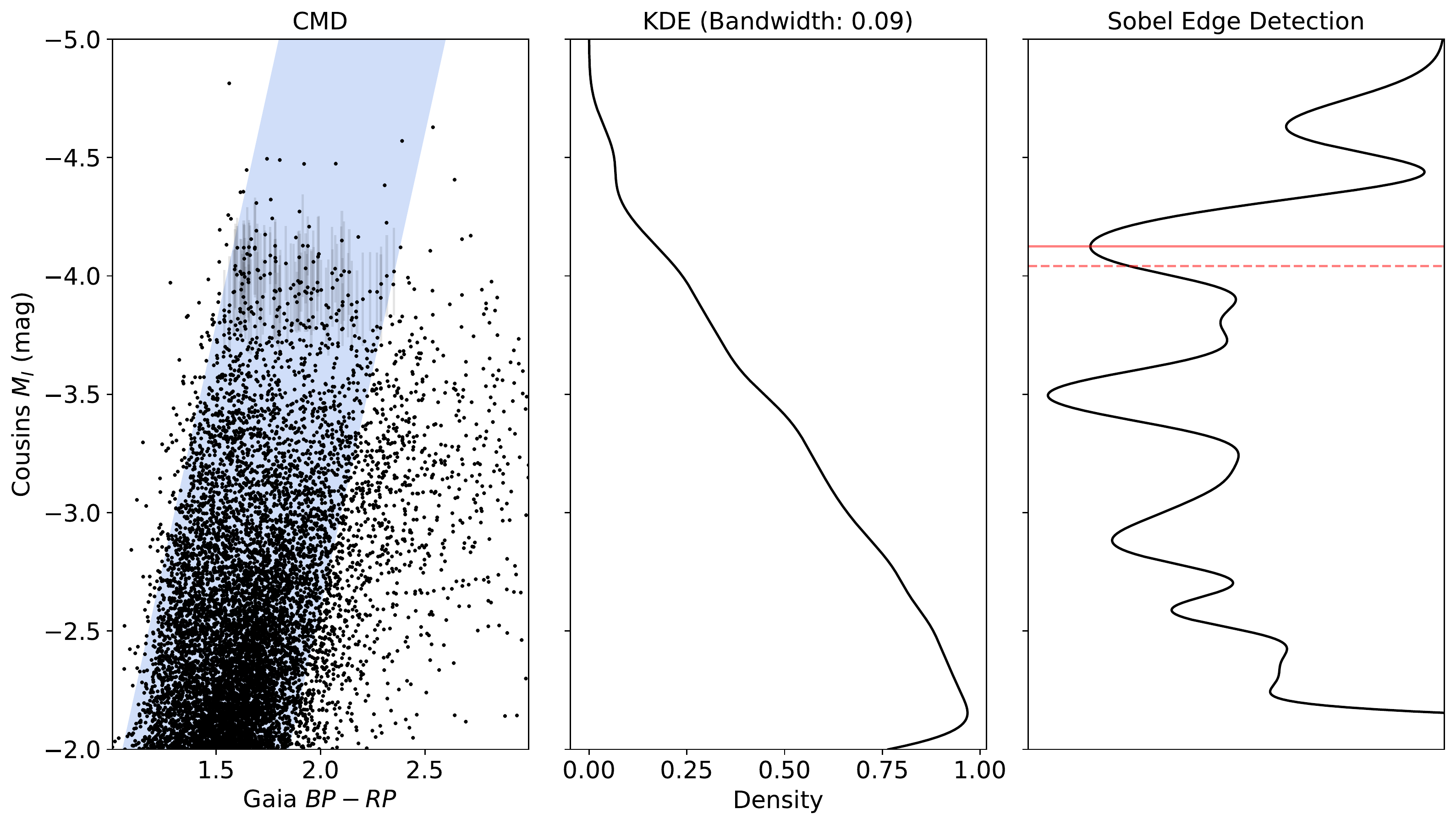}
    \caption{Same as Figure \ref{fig:SKY} when applied to our combined sample containing 9148 stars and an offset correction of 0.082 mag. We find $M_{I}^{TRGB} = -4.042$ mag.}
    \label{fig:combined}
\end{figure*}

As each of our measurements are semi-independent and correlated through the parallaxes, we cannot simply take a weighted average across the four surveys. Instead, we combine the surveys together, forming a sample of 9,148 unique stars. We then obtain an inverse variance weighted mean $I$ for each star and determine $M_{I}^{TRGB} = -4.042$ mag $\pm$0.041 (stat) $\pm$0.031 (sys) (see Figure \ref{fig:combined}). For this approach, we take the notional systematic uncertainty of 0.031 mag by adding the errors in quadrature in Table \ref{Table:systematics}. We also examine the result when `naturally smoothing' our LF using observational uncertainties to obtain the KDE instead of a fixed smoothing kernel. We find $M_{I}^{TRGB} = -4.052$ mag, which is 0.01 mag brighter in comparison, but note that this approach is not the most optimal given our observational scatter.

\subsection{Comparison with recent calibrations of the TRGB}

Our $M_{I}^{TRGB}$ values and uncertainties fall within the range reported in those studies. Therefore, we do not offer a definitive constraint on the peak luminosity of the TRGB, but our value does align more with brighter determinations and is the most precise geometric calibration of the TRGB using Milky Way halo RGB stars to date. We also note that the Milky Way TRGB calibration is consistent across different stellar populations when compared with the LMC, SMC and NGC4258.

Two other recent studies have utilised Gaia parallaxes to achieve this goal. In comparison, we find a significant difference in our work and that of \cite{Li_2022}. They employed a maximum likelihood method, whereas we used Sobel edge detection and also utilised APASS for one of our photometric surveys . We find a brighter TRGB luminosity, which leads to a smaller $H_{0}$ value. We instead implement more specific cuts regarding our colour selection (blue TRGB), extinction and photometric errors. We note that when using the transformations to convert APASS $i$ to Cousins $I$ in \cite{Li_2022}, we obtain $M_{I}^{TRGB} \sim -3.92$ mag. However, we find an offset of 0.05$-$0.10 mag (depending on colour selection) when comparing with the Stetson curated catalogue of standard stars in Cousins $I$ after the transformations. This offset is more significant compared with the other surveys. We instead derive our own colour transformations (see Figures \ref{fig:atlas_offset}, \ref{fig:apass_offset}, \ref{fig:sky_offset}) by comparing each survey with the Stetson curated catalogue \citep{Pancino_2022} and then convert to Cousins $I$.

\cite{Mould_2019} used a chi-squared approach of Milky Way field stars with SkyMapper DR2 and Gaia DR2 parallaxes to calibrate the TRGB zero-point, finding $M_{I}^{TRGB} \sim -4.00 \pm 0.076$ mag. Our results are consistent within errors when using our methodology and the more recent SkyMapper DR3 and Gaia DR3 parallaxes. We were also able to make use of the more precise parallax measurements from Gaia DR3, and can more finely select stars with smaller  uncertainties via our FPC.

\subsection{Limitations and Future Work}

Improving and constraining the measurement of the TRGB requires modeling the distribution of stars in the Milky Way. However, this task becomes increasingly difficult with the presence of AGB stars, which obscure the TRGB transition point. An improvement to our simulations and derived offset values, is to more accurately model AGB stars and constrain the impact of measuring the TRGB. This can be achieved by using up-to-date isochrones with AGB stellar tracks. We run our simulations using a single isochrone to uncover small offsets from the correct TRGB magnitude. Given the halo region contains a mix of stellar populations with a range of metallicities this is an unknown systematic in our approach. We do not know the spread of metallicities from our Gaia sample, but for each star this could be estimated to the first order using the colours and absolute magnitudes. More diverse simulations could be constructed based on the observational metallicities and more accurately model the Milky Way halo stellar population.

We experimented with different selection cuts, which included dropping the Galactic latitude in increments down to ($|b| > 20)$. Although the number of stars increased considerably, by a factor of 2.6, it also caused the scatter in the TRGB to increase. Therefore, we decided to focus on Milky Way halo stars where the effects of crowding, metal-rich disk stars, and dust are less pronounced. Another possibility for improving the selection of halo stars in the Milky Way is to explore alternative regions with low extinction to increase our sample size. However, we were unable to justify the inclusion of these cuts due to the detection of the TRGB scatter being optimal at the higher galactic latitudes. 

While an important limitation  impacting the accuracy of detecting the TRGB is the number of objects in our samples, we find that it is vital to reduce the systematics involved in calibrating out the offset through simulations. The magnitude errors (which are driven by the parallax uncertainties) define the photometric scatter in our simulations and hence in the obtained offset corrections.  An increased sample of RGB stars with high-quality parallaxes will be obtained in the upcoming Gaia DR4 release and decrease statistical errors. Additionally, it is expected that the parallax offset corrections will be more accurately constrained, especially for the brighter, red stars used in calibrating the TRGB. It is also anticipated that the synthetic photometry derived from Gaia DR3 \cite{Gaia_2022_specphot} will be improved in DR4, resulting in an all-sky geometric calibration of higher precision compared to our current analysis.

\subsection{Impact on Cosmology}

The absolute brightness of SNe Ia is needed to determine $H_{0}$, where distances to nearby SN Ia host galaxies are independently measured using standard candles such as Cepheids and the TRGB.  One of the significant advantages of using the TRGB as a standard candle is that it can be found in all types of galaxies. Additionally, when targeting the TRGB in halo regions, the impact of reddening caused by dust and crowding impacts is minimal, unlike in star-forming galaxies that typically host Cepheids. Ultimately, using distances derived using the TRGB can help increase the variety of $H_{0}$ anchor galaxies which can be cross-calibrated with Hubble flow SNe Ia galaxies.

The near-infrared capabilities of JWST will be able to extend TRGB observations and hence the sample of $H_{0}$ anchor galaxies to around 80 Mpc, ultimately helping to reduce the scatter in calibrating SNe Ia. The current scatter for nearby SN Ia host galaxies calibrated using the TRGB is around 0.10 mag \citep{Freedman_2021}. While the number of $H_{0}$ calibrators is increasing, currently $\sim$ 42 \citep{Riess_2021}, the number of new observed SN Ia is limited. JWST will also be able to incorporate the zero-point calibration of the TRGB to determine distances to a number of anchor galaxies which can be measured using surface brightness fluctuations \citep{Blakeslee_2021}. Importantly, this will be a new independent measurement of $H_{0}$, separate from the Cepheid and SN Ia distance ladder, that can better probe systematics in measured distances.

A brighter zero-point calibration translates to a smaller expansion rate, becoming more consistent with early universe measurements, while dimmer values agree with the late universe measurements of a faster expansion rate. Using Equation 4 in \cite{Riess_2016}, we find that an updated TRGB zero-point calibration ($\sim$ 0.05 mag fainter) results in a 1.6 km\,s$^{-1}$ Mpc$^{-1}$ increase in $H_{0}$. Future work will involve measuring $H_{0}$ using our zero-point calibration of the TRGB.
 
\section{Conclusion}\label{Section6}
The TRGB standard candle plays a crucial role in $H_{0}$ measurements as a bridge between Cepheids and SNe Ia.
In this paper, we utilise the latest parallax measurements from Gaia DR3, and four photometric surveys, to obtain a geometric calibration of the TRGB luminosity. Each survey is converted to Cousins I magnitudes and then we have implemented our Sobel edge detection method. 
This introduced a systematic offset tied to photometric errors and  the smoothing of our luminosity function. We show how this bias can be calibrated out by using simulations with PARSEC isochrones and apply the offset correction to our observational data.
We find $M_{I}^{TRGB} = -4.042 \pm 0.041$ (stat) $\pm0.031$ (sys) mag, which is consistent with other recent zero-point calibrations of the TRGB. Our geometric calibration of $M_{I}^{TRGB}$ in the Milky Way is the most accurate to date. However, even greater precision can be achieved in future surveys such as Gaia DR4, which will provide more precise parallax measurements.

%The Acknowledgements section is not numbered. Here you can thank helpful
%colleagues, acknowledge funding agencies, telescopes and facilities used etc.

\section*{Acknowledgements}
We thank an anonymous referee for insightful feedback which improved the quality of the paper.
MD would like to acknowledge support through an Australian
Government Research Training Program Scholarship. This research
was supported by the Australian Research Council The Centre of
Excellence for Dark Matter Particle Physics (CDM; project number
CE200100008) and the Australian Research Council Centre of Excellence for Gravitational Wave Discovery (OzGrav; project number
CE170100004). This project/publication was made possible through
the support of a grant from the John Templeton Foundation. The
authors gratefully acknowledge this grant ID 61807, Two Standard
Models Meet. The opinions expressed in this publication are those
of the author(s) and do not necessarily reflect the views of the John
Templeton Foundation.

This work has made use of data from the European Space Agency (ESA) mission
{\it Gaia} (\url{https://www.cosmos.esa.int/gaia}), processed by the {\it Gaia}
Data Processing and Analysis Consortium (DPAC,
\url{https://www.cosmos.esa.int/web/gaia/dpac/consortium}). Funding for the DPAC
has been provided by national institutions, in particular the institutions
participating in the {\it Gaia} Multilateral Agreement. 

The Pan-STARRS1 Surveys (PS1) and the PS1 public science archive have been made possible through contributions by the Institute for Astronomy, the University of Hawaii, the Pan-STARRS Project Office, the Max-Planck Society and its participating institutes, the Max Planck Institute for Astronomy, Heidelberg and the Max Planck Institute for Extraterrestrial Physics, Garching, The Johns Hopkins University, Durham University, the University of Edinburgh, the Queen's University Belfast, the Harvard-Smithsonian Center for Astrophysics, the Las Cumbres Observatory Global Telescope Network Incorporated, the National Central University of Taiwan, the Space Telescope Science Institute, the National Aeronautics and Space Administration under Grant No. NNX08AR22G issued through the Planetary Science Division of the NASA Science Mission Directorate, the National Science Foundation Grant No. AST-1238877, the University of Maryland, Eotvos Lorand University (ELTE), the Los Alamos National Laboratory, and the Gordon and Betty Moore Foundation.

The national facility capability for SkyMapper has been funded through ARC LIEF grant LE130100104 from the Australian Research Council, awarded to the University of Sydney, the Australian National University, Swinburne University of Technology, the University of Queensland, the University of Western Australia, the University of Melbourne, Curtin University of Technology, Monash University and the Australian Astronomical Observatory. SkyMapper is owned and operated by The Australian National University's Research School of Astronomy and Astrophysics. The survey data were processed and provided by the SkyMapper Team at ANU. The SkyMapper node of the All-Sky Virtual Observatory (ASVO) is hosted at the National Computational Infrastructure (NCI). Development and support of the SkyMapper node of the ASVO has been funded in part by Astronomy Australia Limited (AAL) and the Australian Government through the Commonwealth's Education Investment Fund (EIF) and National Collaborative Research Infrastructure Strategy (NCRIS), particularly the National eResearch Collaboration Tools and Resources (NeCTAR) and the Australian National Data Service Projects (ANDS).

This research was made possible through the use of the AAVSO Photometric All-Sky Survey (APASS), funded by the Robert Martin Ayers Sciences Fund and NSF AST-1412587.

%%%%%%%%%%%%%%%%%%%%%%%%%%%%%%%%%%%%%%%%%%%%%%%%%%
\section*{Data Availability}
Each of the surveys used in our work are publically available at the following sources and/or can be queried directly using TAP services such as TOPCAT.

Gaia: https://gea.esac.esa.int/archive/

SkyMapper: https://skymapper.anu.edu.au/how-to-access/

APASS: https://www.aavso.org/apass

ATLAS: https://archive.stsci.edu/hlsp/atlas-refcat2

%Gaia Synthetic Photometry Catalogue; GSPC; accessible from the table synthetic$_$photometry$_$gspc in the Gaia ESA Archive)


\typeout{}
\bibliographystyle{mnras}
\bibliography{ref} % if your bibtex file is called example.bib

%\begin{thebibliography}{99}
%\end{thebibliography}

%%%%%%%%%%%%%%%%%%%%%%%%%%%%%%%%%%%%%%%%%%%%%%%%%%

%%%%%%%%%%%%%%%%% APPENDICES %%%%%%%%%%%%%%%%%%%%%

\appendix
\section{Extra material}\label{Appendix}

\begin{figure}
    \centering  \includegraphics[scale=0.47]{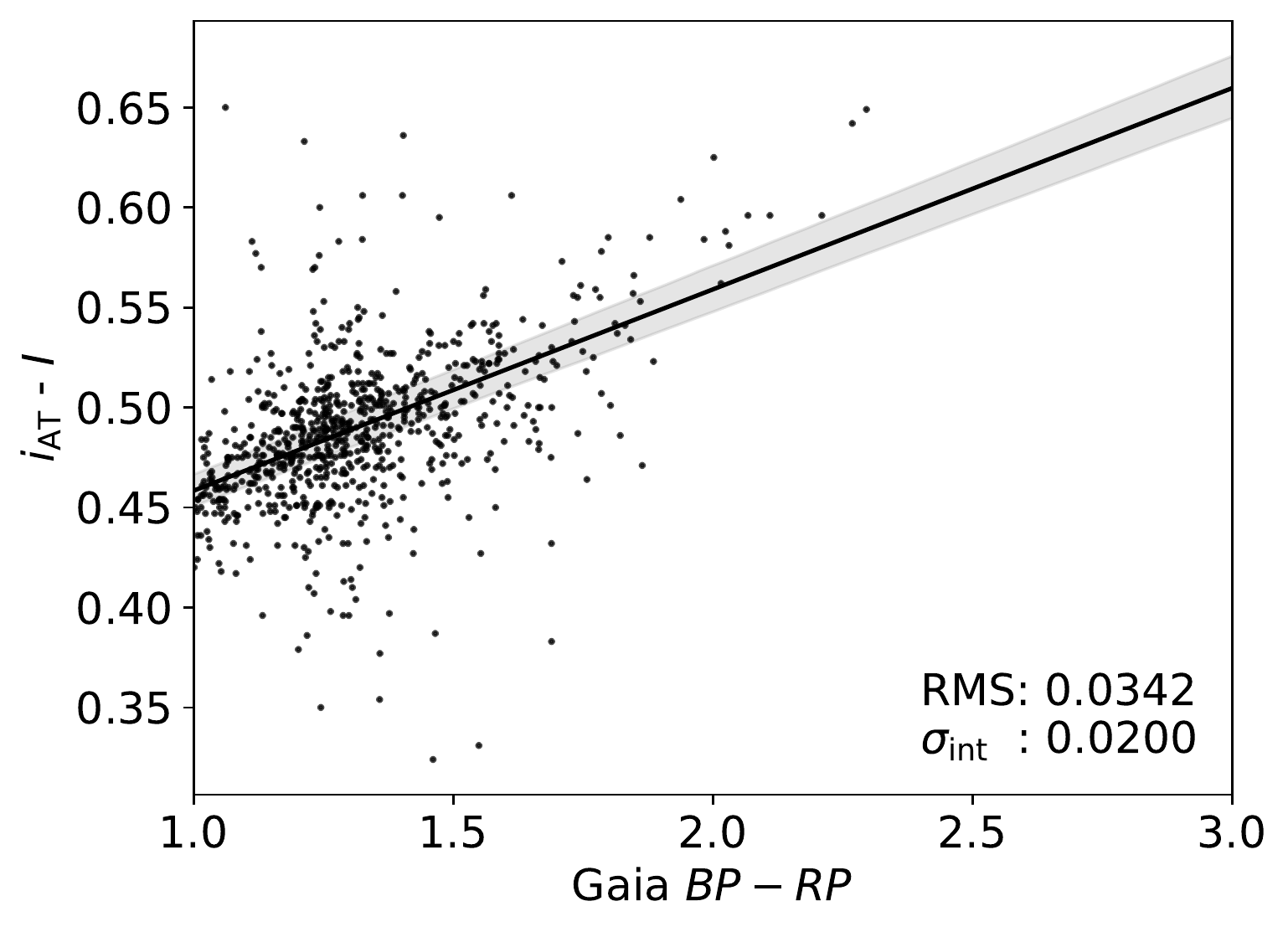}
    \caption{ATLAS: Fitting the relationship between i and Cousins I for objects cross-matched with Gaia DR3 using a collection of Stetson standard stars \citep{Pancino_2022}. The transformation equations are described in Section \ref{transformations}.}
    \label{fig:atlas_offset}
\end{figure}

\begin{figure}
    \centering  \includegraphics[scale=0.47]{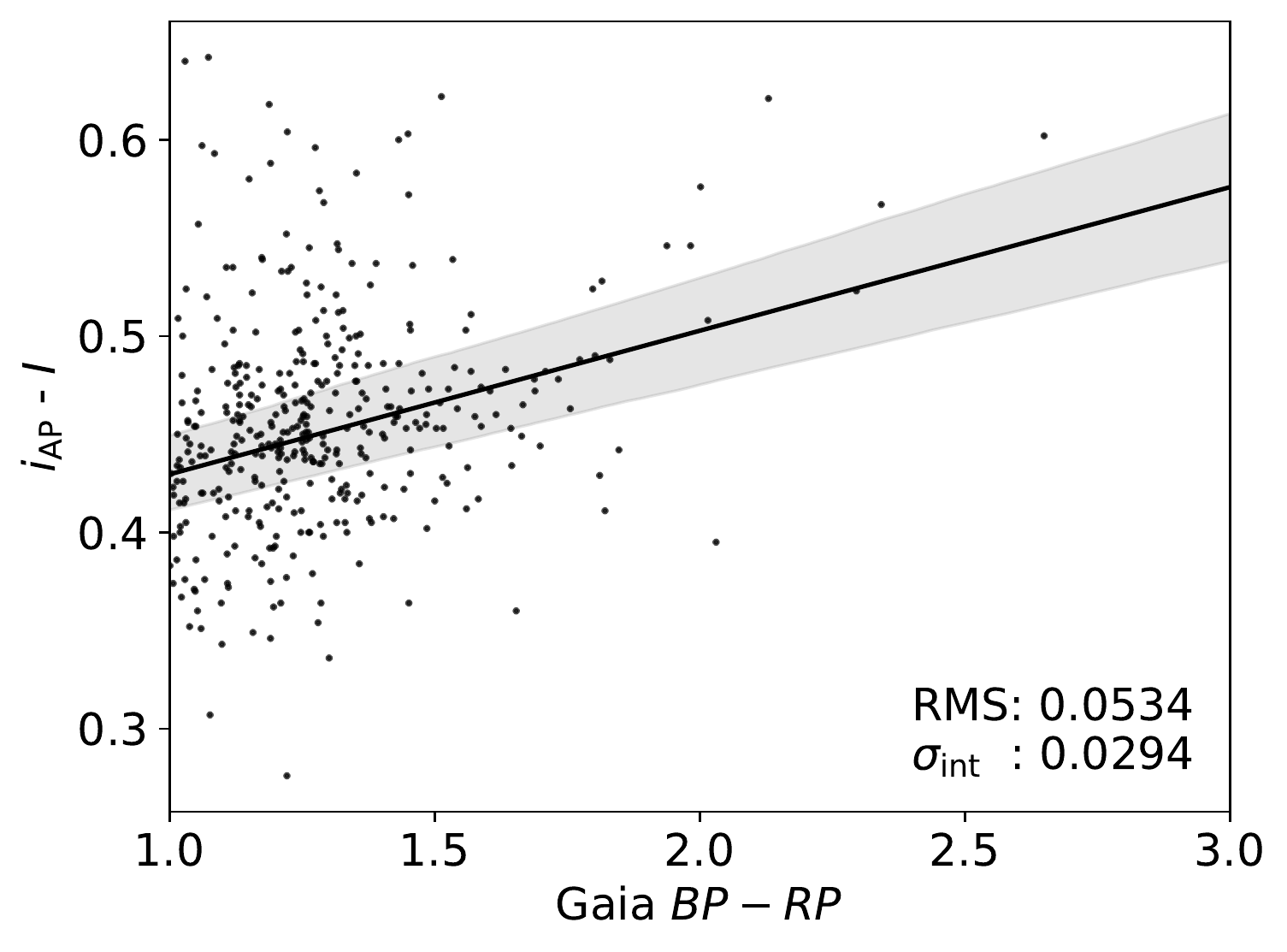}
    \caption{Similar to Figure \ref{fig:atlas_offset}, but for the APASS photometric transformation.}
    \label{fig:apass_offset}
\end{figure}

\begin{figure}
    \centering  \includegraphics[scale=0.47]{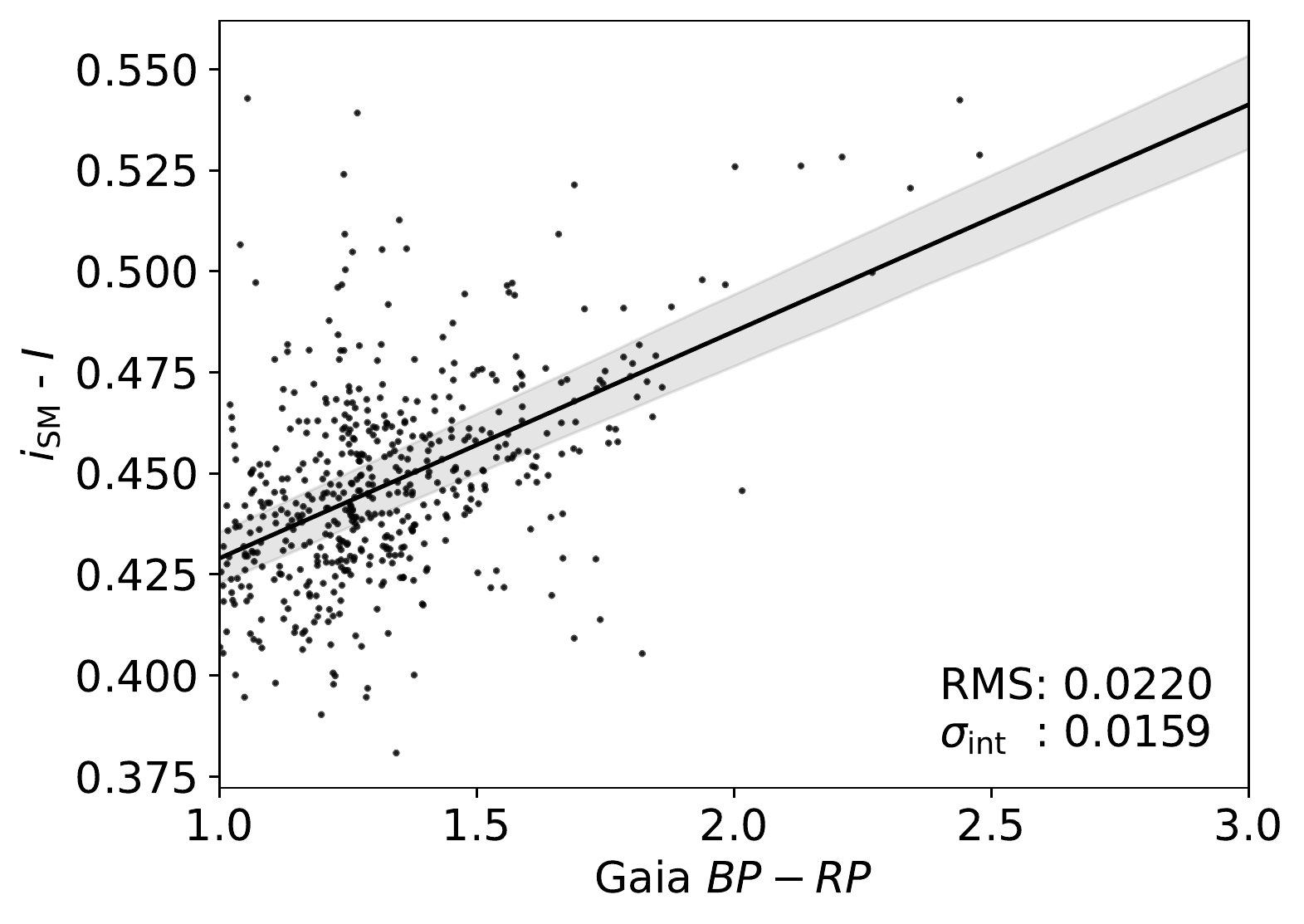}
    \caption{Similar to Figure \ref{fig:atlas_offset}, but for the SkyMapper photometric transformation.}
    \label{fig:sky_offset}
\end{figure}

\begin{figure}
    \centering  \includegraphics[scale=0.47]{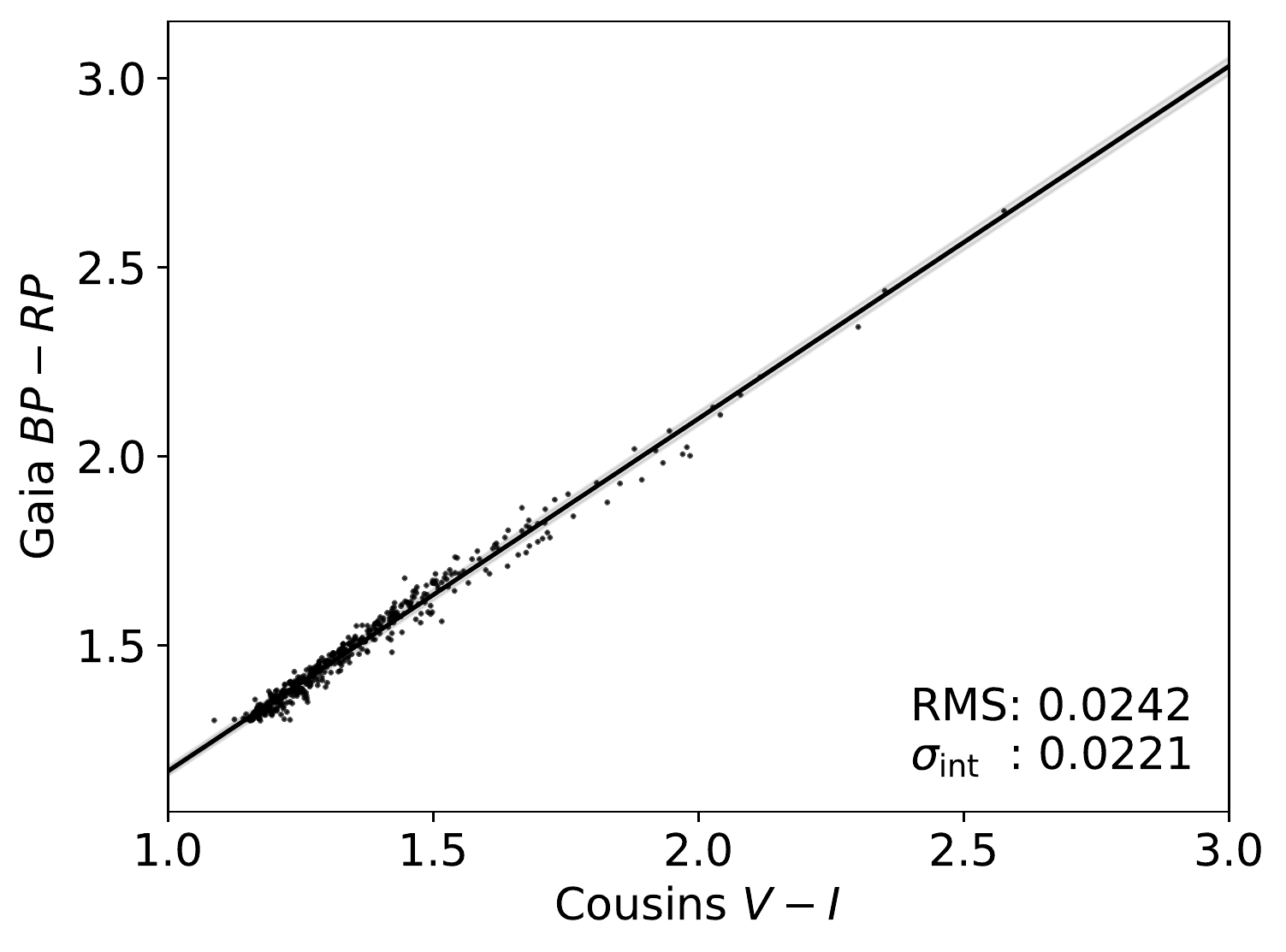}
    \caption{Using the curated catalogue of Stetson standard stars matched with Gaia DR3 \citep{Pancino_2022}, we derive the relationship for RGB stars, where $(BP-RP)$ = 0.932($V-I$) + 0.235.}
    \label{fig:colour}
\end{figure}

%%%%%%%%%%%%%%%%%%%%%%%%%%%%%%%%%%%%%%%%%%%%%%%%%%

% Don't change these lines
\bsp	% typesetting comment
\label{lastpage}
\end{document}